\documentclass[pre,showpacs,twocolumn,groupedaddress,floatfix, nofootinbib, a4paper]{revtex4}
\usepackage[utf8]{inputenc}
\usepackage{epsfig}
\usepackage[T1]{fontenc}
\usepackage{graphicx}
\usepackage{amssymb}
\usepackage{amsmath}
\usepackage{color}
\usepackage{times}
\usepackage{bm}
\usepackage{array}
\usepackage[left=1.5cm,top=2cm, bottom=2cm]{geometry}

\usepackage{mathpazo}
\usepackage{helvet}
\usepackage [euler-digits, euler-hat-accent]{eulervm}
\usepackage [mathscr] {eucal}

\begin{document}

\title{Noise in random Boolean networks}

\author{Tiago P. Peixoto}
\email[]{tiago@fkp.tu-darmstadt.de}
\author{Barbara Drossel}
\email[]{drossel@fkp.tu-darmstadt.de}
\affiliation{Institut für Festkörperphysik, TU Darmstadt, Hochschulstrasse 6,
  64289 Darmstadt, Germany}

\date{\today}

\begin{abstract}
We investigate the effect of noise on Random Boolean Networks. Noise
is implemented as a probability $p$ that a node does not obey its
deterministic update rule.  We define two order parameters, the
long-time average of the Hamming distance between a network with and
without noise, and the average frozenness, which is a measure of the
extent to which a node prefers one of the two Boolean states.  We
evaluate both order parameters as function of the noise strength,
finding a smooth transition from deterministic ($p=0$) to fully
stochastic ($p=1/2$) dynamics for networks with $K\le2$, and a first
order transition at $p=0$ for $K>2$. Most of the results obtained by
computer simulation are also derived analytically. The average Hamming
distance can be evaluated using the annealed approximation. In order
to obtain the distribution of frozenness as function of the noise strength,
more sophisticated self-consistent calculations had to be performed.
This distribution is a collection of delta peaks for $K=1$, and it has
a fractal sructure for $K>1$, approaching a continuous distribution in
the limit $K\gg1$.
\end{abstract}


\maketitle


\section{Introduction}\label{sec:intro}
Random Boolean Networks
(RBNs)~\cite{kauffman_metabolic_1969,drossel_random_2008} have been used as a
simple model for a variety of dynamical systems consisting of interacting units,
such as neural networks~\cite{rosen-zvi_multilayer_2001}, social
networks~\cite{moreira_efficient_2004} and, more prominently, gene regulatory
networks~\cite{lagomarsino_logic_2005, kauffman_metabolic_1969}. RBNs are
composed of Boolean nodes that are coupled to each other. In the case of gene
regulatory networks, the Boolean state is a step-function approach to the
expression level of a particular gene. Despite this loss of detail, the most
important features of gene regulatory processes are still captured in many
cases, since they should not depend on biochemical details, but on the desired
sequence of events in the cell~\cite{bornholdt_systems_2005}.

So far, the dynamics of RBNs have mostly been studied using deterministic update
rules. The dynamics of such models are non-ergodic, with periodic attractor
trajectories in state space. Once the system has reached an attractor, it
remains there. Another important property of RBNs is a phase transition, which
occurs when the number $K$ of inputs per node is changed. For unbiased networks,
the dynamics exhibit a frozen phase at $K=1$, where local perturbations die out
quickly and most attractors are fixed points, and a ``chaotic'' phase at $K>2$,
where perturbations increase exponentially fast and attractors have very long
periods. At the boundary $K=2$ between those two phases are the so-called
``critical'' networks, where perturbations increase algebraically with time.
Originally, it was suggested by Kauffmann~\cite{kauffman_metabolic_1969} that
such critical networks are best suited to model real systems, which are
supposedly poised ``at the edge of chaos''. In the meantime, there is agreement
that RBNs of all three types have only limited validity when applied to real
systems.

Real networks usually have some level of stochastic behaviour, and for this
reason several authors have investigated RBNs under the influence of
stochasticity. For instance, in~\cite{greil_dynamics_2005} the nodes of RBNs
were updated in a completely random order. This update method preserves the
non-ergodicity of the system, and it is still possible to identify distinct
attractors. Attractors are in this case defined as sets of states all of which
are visited for a non-vanishing proportion of time during the same trajectory.
The stochastic update sequence vastly reduces the number of attractors of
critical RBNs, which becomes a power law as function of the network size.
Similar results are obtained when the update sequence deviates only slightly
from a synchronous update~\cite{klemm_stable_2005}.  Such a power law was for a
long time falsely believed to occur in deterministic RBNs
~\cite{kauffman_metabolic_1969,samuelsson_superpolynomial_2003,drossel_number_2005-1,drossel_number_2005}.

Instead of introducing stochasticity into the update times, other authors
introduce it into the update functions.
In~\cite{shmulevich_probabilistic_2002}, probabilistic Boolean functions are
used, where a set of several Boolean functions is assigned to each node, and at
each time step one of these is chosen randomly with a given
probability. According to~\cite{shmulevich_probabilistic_2002}, this model is
more realistic than models with a purely deterministic update scheme.

However, the most important way of introducing noise into a RBN is in
form of a ``temperature'', leading often to ergodic behavior.  The
effect of thermal noise on Ising spins on a network was studied
in~\cite{indekeu_special_2004,aleksiejuk_ferromagnetic_2002}, where a
``ferromagnetic'' transition from the ordered to the disordered phases
was observed at a critical noise strength value.  In the language of
gene regulatory networks, a temperature manifests itself as
fluctuations in the protein concentrations, so that a gene may not
always be turned on or off, given the same expression state of the
other genes~\cite{mcadams_stochastic_1997}. This effect can be
included into models by allowing a deviation from the deterministic
update rule with a certain
probability. In~\cite{fretter_response_2008}, for instance, a subset
of nodes were perturbed in this way (this corresponds to turning on
the temperature for a short time interval), and the response of the
dynamics to this perturbation was evaluated, giving information about
the basin structure of the system.  Miranda et
al.~\cite{miranda_noise_1989} studied the effect of a permanently
acting temperature by introducing a fixed probability $p$ that the
state of a node becomes the opposite of what it should be according to
the deterministic update rule.  They evaluated the average crossing
time between trajectories in state space which started from different
initial states as function of noise strength~\cite{miranda_noise_1989,
golinelli_barrier_1989, qu_numerical_2002}. By sampling the entire
state space of small networks ($N\le20$), it was found that the
``barriers'', which correspond to the attractor basin boundaries, can
be crossed with non-vanishing probability when $p>0$, although the
characteristic times may be large. This means that the system is
always ergodic. This type of noise has also been studied for Boolean
networks with threshold functions, corresponding to a majority update
rule~\cite{huepe_dynamical_2002}. This system undergoes a second order
phase transition at a critical noise strength from an ordered
dynamical phase, where all nodes assume the same value for the
majority of time, to a disordered phase where nodes assume both states
equally often.

In this work, we investigate the effect of ongoing stochastic noise on
RBNs. Following~\cite{miranda_noise_1989, golinelli_barrier_1989,
qu_numerical_2002}, noise strength is tuned via a probability $p$ that
a node does not obey its deterministic update rule. We monitor the
transition from fully deterministic dynamics ($p=0$) to purely
stochastic dynamics ($p=1/2$) as the noise strength is
varied. Differently from~\cite{miranda_noise_1989,
golinelli_barrier_1989, qu_numerical_2002}, we are interested in the
behaviour of the networks in the limit of large system size, where it
is impossible to explore large parts of the state space. In order to
characterize the transition from zero to infinite temperature, we define two
order parameters: the long-time average of the Hamming distance
between a network with and without noise, and the average frozenness,
which is a measure of the extent to which a node prefers one of the
two Boolean states.  We find, both analytically and numerically, that
this transition for the Hamming distance is continuous for $K\le2$,
and discontinuous at $p=0$ for $K>2$, when the Hamming distance is
considered. This distinction is a direct consequence of the phase
transition from frozen to chaotic dynamics in the deterministic
model. The frozenness shows a smooth transition for all values of
$K$. The distribution of frozenness shows a surprising richness in
structure, as revealed by computer simulations. For $K\le2$ and for
$K\gg1$, we succeeded in reproducing this structure as function of $p$
by analytical considerations.

The remainder of this paper is divided into the following parts: In
Sec.~\ref{sec:model} we define the RBN model and the type of noise used for our study.  In
Sec.~\ref{sec:Hamming}, we  define the first order parameter,  the
Hamming distance, and evaluate it numerically
and analytically. In Sec.~\ref{sec:frozenness}, we define the second
order parameter, the frozenness, and evaluate it using computer
simulations and analytical considerations. 
 Finally, we summarize and discuss our findings  in Sec.~\ref{sec:conclusion}.

\section{Model}\label{sec:model}

A Boolean network is defined as a directed network of $N$ nodes representing
Boolean variables $\bm{\sigma} \in \{1,0\}^N$, which are subject to a
dynamical update rule,
\begin{equation}
  \bm{\sigma}(t+1) = \bm{f}\left(\bm{\sigma}(t)\right),
\end{equation}
where $f_i$ is a function assigned to node $i$ that depends
exclusively on the states of its inputs. 

We introduce noise into the system through a probability $p$ that a
node does not obey its deterministic update rule,
\begin{equation}\label{eq:bnoise}
  \bm{\sigma}(t+1) = \bm{f}\left(\bm{\sigma}(t)\right)\veebar\bm{n},
\end{equation}
where $\bm{n}$ is a random vector, with elements $n_i$ being $1$ with
probability $p$ and $0$ otherwise. The symbol $\veebar$ represents the
``exclusive or'' Boolean operation. Hence, for $p=0$ the deterministic behaviour
is recovered, and for $p=1/2$ the dynamics is completely stochastic.

RBNs are a special case of Boolean networks, where all possible
Boolean functions are assigned randomly to each node with the same
probability, and where the nodes are randomly connected.  The number
of inputs of each node is fixed at a value $K$. The random wiring
leads to a Poisson distribution with mean $K$ for the number of
outputs. When updated deterministically, RBNs are in the frozen phase
for $K=1$. After a transient time, they reach an attractor
where all nodes (or all nodes apart from a small number) are
permanently frozen in one of the two Boolean states. Networks with
larger $K$ have also a \emph{frozen core} of nodes for $p=0$, and the
nodes belonging to it become frozen after a transient time. For $K=2$, all but of the order
of $N^{2/3}$ nodes belong to the frozen core. With increasing $K$, the
frozen core contains an ever smaller proportion of nodes. For $K=2$,
the nonfrozen part of the network consists of several independent
components. Each of these components contains a set of \emph{relevant
nodes}, which are connected such that there is at least one
feedback loop among them, and ``trees'' of nonfrozen nodes which are
rooted in the relevant nodes and which are slaved to the dynamics of
the relevant nodes.

\section{Average Hamming distance}\label{sec:Hamming}

\subsection{Definition}\label{sec:defHamming}

We use the average in time of the Hamming distance between the states
of two copies of a network in order to quantify the effect of noise on
the dynamics. Consider a given network in the initial state
$\bm{\sigma}(t=0)$, and an exact replica, which is initially in the
same state, $\bm{\sigma}'(t=0)= \bm{\sigma}(t=0)$.  The dynamics of
both networks are evolved in parallel, but noise is applied only to
$\bm{\sigma}'(t)$, as in Eq.~(\ref{eq:bnoise}). The mean  Hamming distance
$h(t)$ between the two networks is defined as 
\begin{equation}\label{eq:Hamming1}
  h(t) =  \frac{1}{N} \langle \left|\sigma_i(t) -
  \sigma_i'(t)\right|\rangle\, ,
\end{equation}
where $\langle \dots \rangle$ denotes the average over the noise. 

The long-time average  $h$ of the Hamming distance is defined as
\begin{equation}\label{eq:Hamming2}
  h = \lim_{T\to\infty}\frac{1}{T}\frac{1}{N} \sum_{t,i} \left|\sigma_i(t) - \sigma_i'(t)\right|.
\end{equation}

If the trajectories become completely uncorrelated after some time, we
have  $h=1/2$. If the trajectories remain closer in state space,
we have $h<1/2$. The case $h>1/2$ does not occur in our model and is
therefore not considered in this paper. 

\subsection{Annealed approximation}

We will first evaluate analytically the Hamming distance by using the so-called
annealed approximation~\cite{derrida_random_1986}. This is a mean field theory,
which neglects correlations between nodes and the finite size of the
network. The annealed approximation corresponds to the behaviour of a
(infinitely large) network where all the edges are randomly rewired at each time
step. Within the annealed approximation, the dynamics of a RBN without noise
(i.e. for $p=0$) is fully specified by the parameter $\lambda$, which is $K$
times the probability that a node changes its state when one (or more) of its
inputs is flipped. For RBNs, we have $\lambda=K/2$, since for any input
combination, there is an equal probability that the output of a function will be
either $0$ or $1$.  When considering two replicas of a network, $\lambda$ is
identical to the mean number of nodes that assume a different state in the two
networks at time $t=1$ when at time $t=0$ the state of only one node was
different.

\begin{figure}[h]
\includegraphics*[width=\columnwidth]{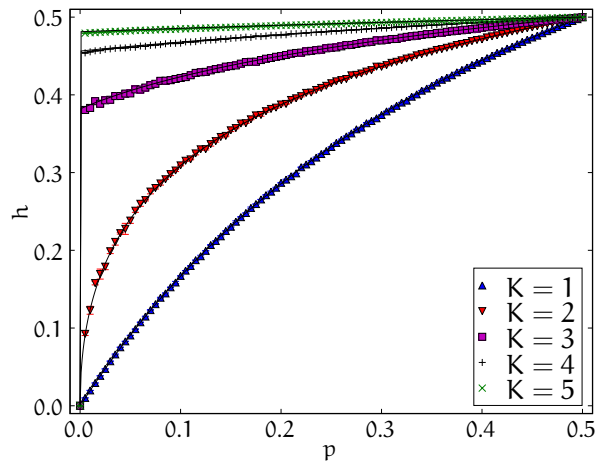}
\caption{(Color online) Average Hamming distance as function of noise strength for
  RBNs of size $N=10^4$ for different values of $K$. Each point was
  obtained by averaging the results over 3 different network
  realizations. The solid lines are the respective steady-state
  solutions of Eq.~(\ref{eq:na}).\label{fig:Hamming-noise}}
\end{figure}

At any time, the Hamming distance between a network with noise and its
twin noiseless counterpart, as described by Eq.~(\ref{eq:Hamming1}),
is simply the fraction of nodes which were changed by noise or by the
effect of previously changed nodes. The time evolution of $h(t)$ can
then be described as the evolution of the population of flipped nodes.
(A node in the replica with noise is called ``flipped'' it its state
deviates from the state it has in the replica without noise.)  Let
$q(h(t))=1-(1-h(t))^K$ denote the probability that a node has at least
one flipped input. Then the probability $h(t+1)$ that a node is
flipped at time $t+1$ can be written as 
\begin{equation}\label{eq:na}
  \begin{split}
    h(t+1) &= \frac{\lambda}{K}q(1-p) + \left(1-\frac{\lambda}{K}\right)pq + (1-q)p\\
           &= \frac{\lambda(1-2p)}{K}\left[1-\left(1-h(t)\right)^K\right] + p, \\
  \end{split}
\end{equation}
where the first term in the first line corresponds to the proportion
of nodes that are flipped by previously flipped nodes (and are not
flipped back by noise), and the second and third term are the
proportion of nodes that are flipped by noise (with or without inputs
being flipped). The fixed point of Eq.~(\ref{eq:na}) determines the
order parameter $h$, for given $K$ and $p$. We evaluated this fixed
point numerically.  Fig.~\ref{fig:Hamming-noise} shows $h$ as function
of the noise strength $p$ for several values of $K$. The solid lines
are the fixed point solutions of Eq.~(\ref{eq:na}), the symbols
represent the result of computer simulations of quenched RBNs. The
agreement between the annealed approximation and the real networks is
very good.

The most striking feature of Fig.~\ref{fig:Hamming-noise} is the existence of a
first-order transition at $p=0$ for $K>2$. This is due to the phase
transition to ``chaotic'' behaviour for $K>2$. In chaotic networks,
even the smallest local perturbations have a global effect. 

\subsection{The Hamming distance on subsets of nodes}

\begin{figure}[h]
\includegraphics*[width=\columnwidth]{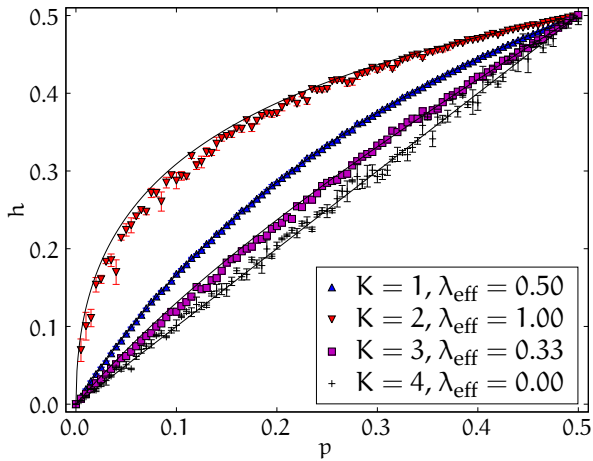}
\caption{(Color online) Average Hamming distance as function of noise for RBN of size $N=10^4$
  for different values of $K$. Only nodes belonging to the frozen core of the
  network (without noise) were considered. Each point on each curve was obtained
  by averaging the results for 3 different network realizations. The solid lines
  are the respective solutions of Eq.~(\ref{eq:na}), with $\lambda_{\text{eff}}$
  being used as a fit parameter. \label{fig:Hamming-fr}}
\end{figure}

We next evaluate separately the Hamming distance
for the frozen core and for the nonfrozen part of the network. 
Fig.~\ref{fig:Hamming-fr} shows the long-time Hamming distance,
evaluated only for the nodes that belong to the frozen core.  These
curves can be fitted using the annealed approximation
Eq.~(\ref{eq:na}) under the condition that the factor $\lambda/K$ on the
right-hand side of Eq.~(\ref{eq:na}) (representing the probability
that a node is flipped when at least one input is flipped) is replaced
with $\lambda_{\text{eff}}/K$, with $\lambda_{\text{eff}}$ being used as a fit
parameter. For $K=1$ and $2$, the frozen core is virtually
indistinguishable from the rest of the network, and
$\lambda_{\text{eff}}=\lambda$, but for $K>2$, $\lambda_{\text{eff}}$
decreases with increasing $K$. The reason is that 
the frozen core becomes composed mainly of nodes with constant functions
which have $\lambda_{\text{eff}}=0 $.

Before evaluating $h$ for the nonfrozen nodes, let us consider the simplest
possible connected set of nonfrozen nodes, which is a simple loop. For $K=1$ and
$K=2$, such simple loops of nonfrozen nodes play an important role at
determining the attractors with periods larger than
1~\cite{kaufman_properties_2005}, however, a considerable fraction of $K=2$
networks also have more complex relevant components.  The effect of noise on
such loops is very different from its effect on the frozen core, since if one of
its nodes is flipped, this flip propagates indefinitely around the loop.  One
can calculate the accumulation of flips on such loops by considering the average
Hamming distance at a given time between a loop without and with noise,
\begin{equation}
\begin{split}\label{eq:Hamming-loop}
h(t) &= \sum_{i=0}^{\lfloor \frac{t-1}{2}\rfloor}{t \choose 2i+1}p^{2i+1}(1-p)^{t-(2i+1)}\\
     &\approx \sum_{i=0}^{\lfloor  \frac{t-1}{2}\rfloor} \frac{(tp)^{2i+1} e^{-tp}}{(2i+1)!}\\
     &\approx \frac{1}{2}\left(1-e^{-2tp}\right)\, .
\end{split}
\end{equation}
The first equation evaluates the probability that a node has been
flipped an odd number of times, and the subsequent transformations 
are valid for $t \gg 1$. The Hamming distance approaches the value
$1/2$ with an exponential decay, and with an characteristic time $\tau
= 1/(2p)$. 

\begin{figure}[h]
\includegraphics*[width=\columnwidth]{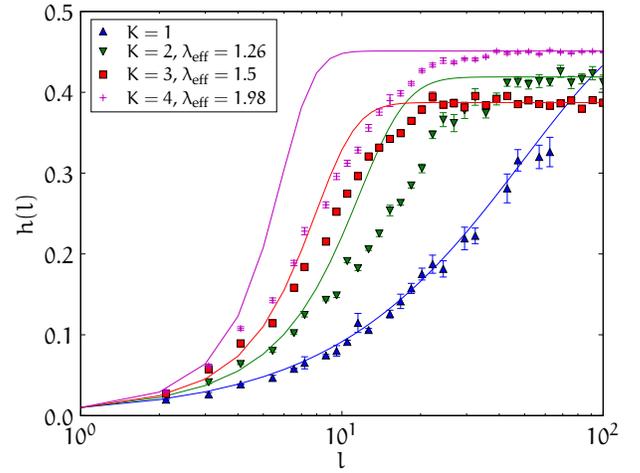}
\caption{(Color online) Average Hamming distance $h(l)$ of the relevant components of RBNs with
  different values of $K$, after a full period $l$, with $p=0.01$. The curves
  were obtained by sampling at least $2\times10^4$ attractors of several
  distinct RBNs of sizes $N=10^2$, $50$ and $25$ for $K\leq2$, $3$ and $4$,
  respectively. The solid lines are given by Eq.~\ref{eq:Hamming-loop} for $K>1$
  and Eq.~\ref{eq:na} for $K\ge2$.\label{fig:Hamming-robust}}
\end{figure}

We evaluated how fast a trajectory leaves an attractor in the presence
of noise by first letting the system approach an attractor and by then
turning on the noise and measuring the Hamming distance $h(l)$ to the
initial state after one attractor period $l$. This is identical to the
distance from the state of the noiseless replica, which returns to the
initial state at time $l$.  Fig.~\ref{fig:Hamming-robust} shows the
values of $h(l)$ for RBNs with different values of $K$.  For $K=1$,
the data match Eq.~(\ref{eq:Hamming-loop}) very well, since the
nonfrozen part of the network in this case can only be composed of
simple loops. For $K\ge2$, the data points are considerably above this
exponential curve because a node can become flipped via many different
paths. The data are better fitted using Eq.~(\ref{eq:na}), in
particular for long periods (i.e. large times). Just as for the case
of the frozen core, $\lambda_{\text{eff}}$ was used as a fit
parameter.

For smaller values of the attractor period, the data are
considerably below the fitted line. The reason is that these attractor
periods are much smaller than typical attractor periods, and
networks with such short attractors are not characteristic of the
ensemble, but have a state-space structure with a smaller set of
recurrent states. Consequently trajectories diverge less fast than
in typical networks. 

\section{Frozenness}\label{sec:frozenness}

\subsection{Definition}

The ``frozenness'' of a network measures the extent to which the nodes
spend more time in one of the two Boolean states. It is zero, when the
nodes spend the same time in both states, and it is 1 when the network
is frozen. The frozenness of node $i$ is defined by the expression
\begin{equation}\label{eq:frozenness}
  \Omega_i =  \left(q_0^{(i)} - q_1^{(i)}\right)^2,
\end{equation}
where $q_\sigma^{(i)}$ is the proportion of time node $i$ is in state $\sigma$,
\begin{equation}
  q^{(i)}_{\sigma} = \lim_{T\to\infty}\frac{1}{T}\sum_{t=0}^T\delta_{\sigma_i(t),\sigma}.
\end{equation}
By eliminating one of the two probabilities from
Eq.~(\ref{eq:frozenness}), we obtain
\begin{equation}
  \Omega_i = \left(2q_{\sigma}^{(i)} - 1\right)^2,
\end{equation}
where $\sigma$ is either $0$ or $1$. 

\begin{figure}[h]
\includegraphics*[width=\columnwidth]{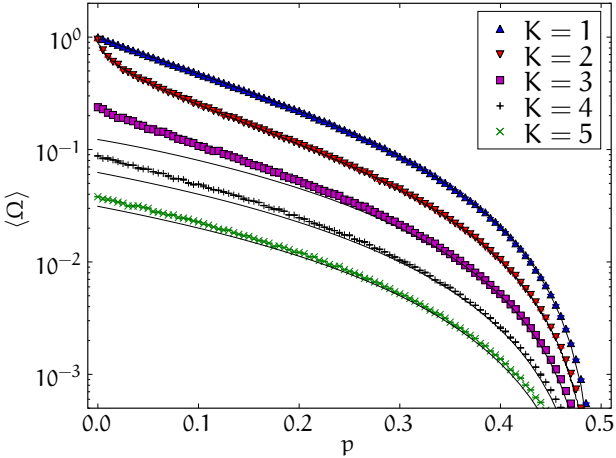}
\caption{(Color online) Frozenness as function of noise strength for RBNs of size $N=10^4$ for
  different values of $K$. Each curve was obtained by averaging the results for
  100 different network realizations. The solid lines correspond to the averages
  of $\rho_\Omega(\Omega|K,p)$, obtained in Secs.~\ref{sec:k1}, \ref{sec:k2}
  and \ref{sec:kg2}, for $K=1$, $2$ and $\ge 3$,
  respectively. \label{fig:frozenness-vs-noise-better}}
\end{figure}

The frozenness $\langle \Omega \rangle$ of the network is obtained by
averaging over the nodes. 
Figure \ref{fig:frozenness-vs-noise-better} shows the frozenness
$\langle \Omega \rangle$ as function of the noise strength $p$
obtained by computer simulations of networks of size $10^4$, for
different $K$.

At $p=0$, the frozenness corresponds obviously to the size of the
frozen core, and it decreases towards 0 as the noise strength
approaches the value 0.5.

In order to derive these curves analytically, the annealed approximation is of
no use, since a network that is rewired during the course of time has very small
frozenness, which is due uniquely to the constant functions. Therefore, a simple
analytical calculation, which does not require the consideration of correlations
between nodes, can only be performed for nodes with constant functions: for such
nodes the frozenness is given by
\begin{equation}
\Omega_i=(1-2p)^2\, .
\end{equation}

In the following, we will present more advanced analytical evaluations
and further computer simulations for RBNs with different values of $K$.

\subsection{$K=1$}\label{sec:k1}

For $K=1$ there are only $2^{2^{1}}=4$ possible Boolean functions, two
of which are constant ($1$ or $0$), and the remaining ones are the
copy ($f(\sigma)=\sigma$) and invert ($f(\sigma)=\lnot \sigma$) functions. As
far as the analysis of frozenness is concerned, there are only two
distinct functions, constant and non-constant, since the output value is not
relevant, but only how often it changes.  Since each of the two types
of functions occurs equally often in a $K=1$ RBN, we have
$\lambda=1/2$, but $K=1$ networks with other values of $\lambda$ can also be constructed.

In a network with $K=1$, each node has only one input, and this input
node also has one input, etc. In order to
evaluate the probability that a node is flipped, one only needs to
consider the chain of those nodes that can have an influence on the
considered node. Nodes with constant functions present a barrier to
the propagation of a perturbation, since they do not respond to a
change in their inputs, and therefore the chain ends (or, more
precisely, begins) at a node with a constant function. 

Without loss of
generality, we define $q^{(i)}$ as being the proportion of time 
node $i$ assumes its most frequent value,
\begin{equation}\label{eq:qdef_k1}
  q^{(i)} \equiv \max(q^{(i)}_\sigma, 1-q^{(i)}_\sigma) \; \in \left[1/2,1\right].
\end{equation}
Since the value of $q^{(i)}$ is fully determined by the distance of
node $i$ to a node with a constant function, we choose the label $i$ in
the remainder of this subsection to signify this distance. If the node
itself has a constant function, we have $i=0$, if the node has a
non-constant function, but its input has a
constant function, we have $i=1$, etc. 

The value of $q^{(0)}$, i.e. for nodes with constant functions,  is simply
\begin{equation}\label{eq:p0_k1}
  q^{(0)}(p) = 1-p.
\end{equation}
For larger values of $i$, we have the recursion relation
\begin{equation}\label{eq:qi_k1}
  \begin{split}
     q^{(i)} &= (1-p)q^{(i-1)} + (1-q^{(i-1)})p\\
           &= \frac{1}{2}(1-2p)^{i+1} + \frac{1}{2},
  \end{split}
\end{equation}
where the solution of the recursion relation was obtained 
using Eq.~(\ref{eq:p0_k1}). The probability of finding a given
$q^{(i)}$ in the network is 
\begin{equation}\label{eq:pq_k1}
  p_q(q^{(i)}|p) = (1-\lambda)\lambda^i = \left(\frac 1 2 \right)^{i+1}\, .
\end{equation}

The frozenness of the network is thus given by
\begin{equation}
\left<\Omega\right>(p)  = \sum_i  p_q(q^{(i)}|p) (1-2 q^{(i)} )^2\, ,
\end{equation}
which is plotted in Figure \ref{fig:frozenness-vs-noise-better} 
and fits the curve for $K=1$ well.

Although networks with $K=1$ have only a discrete set of possible
$q$-values, the distribution of $q$ values appears as a continuum when determined by computer
simulations. There are two reasons for this. First, the data points
for $q$ close to 1/2 (i.e. for $\Omega$ close to 0) are so close to
each other that they cannot be resolved, since a computer simulation
uses a non-vanishing bin size. 
Solving Eq.~(\ref{eq:qi_k1}) for $i$, inserting the result in
(\ref{eq:pq_k1}), and using the relation $\rho_q(q|K=1,p)dq
= p_q(q|K=1,p)$ with $dq=q^{(i)}-q^{(i+1)}$,  we obtain
\begin{equation}
  \rho_q(q|K=1,p) \propto
  (2q-1)^{\frac{\ln\lambda}{\ln(1-2p)}-1}\, ,
\end{equation}
with $\lambda=1/2$ for RBNs.
Thus, in the limit $q\to 1/2$ the distribution of
$2q-1$ follows as a power-law with an exponent given by the above
expression. 
The probability density of $\Omega$  also decays as a power
law, in the limit $\Omega\to 0$, but with a different exponent, since
\begin{equation}\label{eq:o_k1}
  \begin{split}
  \rho_\Omega(\Omega|K=1,p) =&\; \rho_q\left(\left.\frac{\sqrt{\Omega}+1}{2}\right|K=1,p\right)\frac{dq}{d\Omega}\\
                              \sim&\;
                              \Omega^{\frac{\ln\lambda}{\ln(1-2p)}-3/2}\, .
  \end{split}
\end{equation}

Second, a computer simulation averages
only over a finite amount of time, $T$, and therefore the measured values $q'$
are Gaussian distributed around the exact value $q$,  
\begin{equation}\label{eq:q_T}
\begin{split}
  \tilde\rho_{q'}(q'|K=1,T,q) \approx \frac{1}{\sqrt{2\pi q/T}} e^{-\frac{(q'-q)^2}{2q/T}},
\end{split}
\end{equation}

This dependence on $T$  can  be included in
Eq.~(\ref{eq:pq_k1}) to obtain the probability density function for $q$
\begin{equation}
  \begin{split}
    \rho_q(q|K=1,p,T) =&\, (1-\lambda)\sum_{i=0}^\infty \lambda^i \tilde\rho_{q'}(q|T,q^{(i)}).
  \end{split}
\end{equation}
For the distribution of $\Omega$ values, we obtain
\begin{equation}\label{eq:pf_k}
  \rho_\Omega(\Omega|K,p,T) =
  \rho_q\left(\left.\frac{\sqrt{\Omega}+1}{2}\right|K,p,T\right)
  \frac{1}{2\sqrt{\Omega}}.
\end{equation}

\begin{figure}[t]
\includegraphics*[width=\columnwidth]{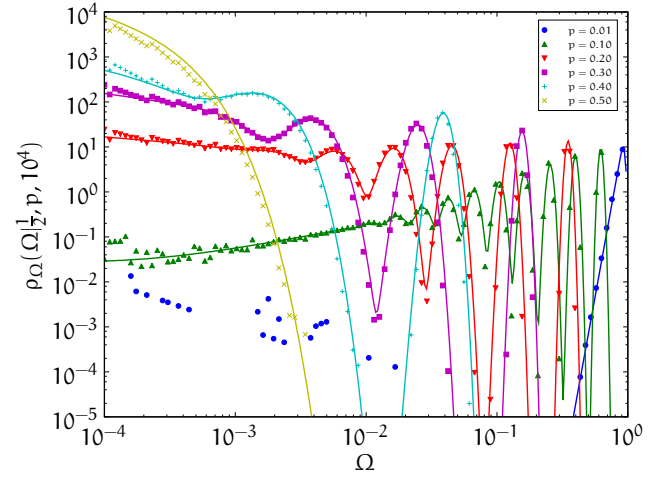}
\caption{(Color online) Distribution of the frozenness for different values of the noise
  strength for RBNs of size $N=10^5$ for $K=1$ and $T=10^4$. Each curve was
  obtained by averaging over 100 different network realizations. The solid lines
  are given by Eq.~(\ref{eq:pf_k}).\label{fig:frozenness-dist-k1}}
\end{figure}

Fig.~\ref{fig:frozenness-dist-k1} shows the distribution of the
frozenness for a quenched network with $K=1$, for several values of
$p$. It can be seen that there is very good agreement with
Eq.~(\ref{eq:pf_k}). The presence of fluctuations significantly
deviates some of the distributions from the expected power-law
decay. For $p=0.01$, the small values of frozenness, which are not in
agreement with the theoretical result, are due to the existence of
loops, which are omitted in the analysis above. The probability that a
node is part of a nonfrozen loop tends to zero as the network becomes
larger, and therefore these points vanish in the limit of infinite
system size.

\subsection{$K=2$}\label{sec:k2}

As $K$ becomes larger, the number of possible functions grows very
fast as $2^{2^K}$, and a detailed analysis of the frozenness, as was
done for $K=1$, becomes more complicated. The values of $q$ are still
discontinuously distributed, but their number increases fast with $K$,
due to the numerous combinations of Boolean functions that can
determine the $q$-values of the $K$ inputs of a node and thus, in
combination with the node's Boolean function, the $q$-value of its
output.
Here
we will lay out the basic considerations needed to obtain the
distribution of $q$ for all $K>1$, and we will obtain by numerical
iteration the distribution for $K=2$. Without loss of generality, we
redefine $q$ as $q \equiv q_{\sigma=1}$, i.e., the fraction of
time a given node has the value 1 (as opposed to Eq.~(\ref{eq:qdef_k1}),
which simplified the case $K=1$).

In general, the probability density function $\rho_q(q|K=1,p)$ needs to
account for all possible recursive combinations of output functions
and their inputs. We can thus write the following self-consistent
expression,
\begin{widetext}
\begin{equation}\label{eq:pq_allk}
  \rho_q(q|K,p) = \int_0^1\dotsi\int_0^1 
                  \sum_fp_f\delta(q^{(f)} - q)
                   \prod_{i=1}^K\rho_q(q^{(i)}|K,p)dq^{(i)},
\end{equation}
\end{widetext}
where the sum is taken over all Boolean functions; $p_f$ is the
probability of the $f$th Boolean function ($p_f=2^{-2^k}$ ),
and
\begin{equation}
  q^{(f)} = (1-2p)q^{(f)}(\{q^{(i)}\})+p,
\end{equation}

where $q^{(f)}(\{q^{(i)}\})$ is the value of $q$ for a specific function $f$,
given the values $\{q^{(i)}\}$ of its inputs, for $i = 1,\dots,K$. 

Since Eq.~(\ref{eq:pq_allk}) involves an expression $q^{(f)}(\{q^{(i)}\})$
for all Boolean functions, a general closed solution becomes
unfeasible. However, for $K=2$ Eq.~(\ref{eq:pq_allk}) can at least be
solved numerically, since there are only $16$ possible functions, 
given in Table~\ref{table:qf_k2}. 
Eq.~(\ref{eq:pq_allk}) is then solved by iteration, until convergence
to a self-consistent $q$ distribution is obtained. 
We started with the initial distribution
\begin{equation}
  \rho^0_q(q|K,p) = \frac{1}{2}\delta(q-p) +
  \frac{1}{2}\delta(q-(1-p))\, .
\end{equation}
In the end, we determined the final distribution
$\rho_\Omega(\Omega|K=2,p)$  by using Eq.~(\ref{eq:pf_k}).

\begin{table}
  \begin{tabular}{r|l|l}
    \hline\hline
      $f$ & $f_i(\sigma_1,\sigma_2)$        & $q^{(f)}(q_1,q_2)$ \\ \hline
      $0$ & $0$                             & $0$                 \\
      $1$ & $\sigma_1 \land \sigma_2$       & $q_1q_2$             \\
      $2$ & $\sigma_1 \land \lnot\sigma_2$  & $q_1(1-q_2)$         \\
      $3$ & $\sigma_1$                      & $q_1$               \\
      $4$ & $\lnot\sigma_1 \land \sigma_2$  & $(1-q_1)q_2$         \\
      $5$ & $\sigma_2$                      & $q_2$                \\
      $6$ & $\sigma_1\veebar\sigma_2$       & $q_1 + q_2 - 2q_1q_2$ \\
      $7$ & $\sigma_1\lor\sigma_2$          & $q_1 + q_2 - q_1q_2$  \\
      $8$ & $\lnot(\sigma_1\lor\sigma_2)$   & $1-q^{(7)}(q_1,q_2)$  \\
      $9$ & $\lnot(\sigma_1\veebar\sigma_2)$ & $1-q^{(6)}(q_1,q_2)$ \\
     $10$ & $\lnot\sigma_2$                 & $1-q_2$             \\
     $11$ & $\sigma_1\lor\lnot\sigma_2$     & $q^{(7)}(q_1,1-q_2)$  \\
     $12$ & $\lnot\sigma_1$                 & $1-q_1$             \\
     $13$ & $\lnot\sigma_1\lor\sigma_2$      & $q^{(7)}(1-q_1,q_2)$ \\
     $14$ & $\lnot(\sigma_1\land\sigma_2)$   & $1-q_1q_2$          \\
     $15$ & $1$                            & $1$ \\ \hline \hline
    \end{tabular}
    \caption{Expressions of $q^{(f)}(q_1,q_2)$ for all Boolean functions for
        $K=2$. The Boolean expressions of each function is also given for
        reference.}\label{table:qf_k2}
\end{table}

\begin{figure}[h]
\includegraphics*[width=\columnwidth]{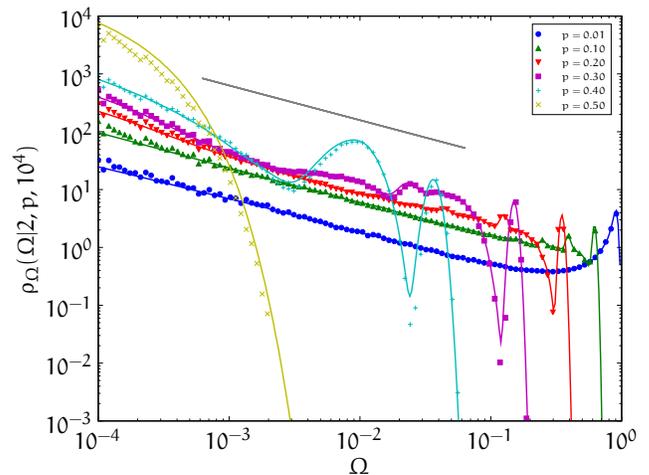}
\caption{(Color online) Distribution of the frozenness of nodes for different noise strength for RBNs of
  size $N=10^5$ and for $K=2$ and $T=10^4$. Each curve was obtained by
  averaging over 100 different network realizations. The solid lines show the
  values of $\rho_\Omega(\Omega|K=2,p,10^4)$ according to Eq.~(\ref{eq:pf_k}). The
  line segment corresponds to a power law with an exponent
  $0.6$.\label{fig:frozenness-dist-k2}}
\end{figure}

\begin{figure}[h]
\includegraphics*[width=\columnwidth]{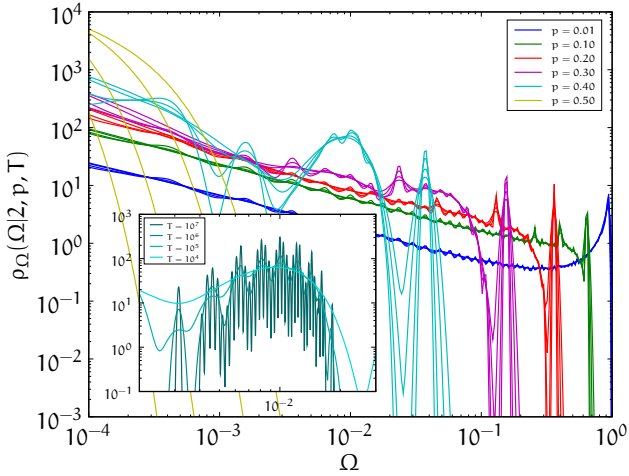}
\caption{(Color online) Expected distribution of frozenness of each node as function of noise
  for RBNs with $K=2$ and different values of $T$. The curves represent
  Eq.~(\ref{eq:pf_k}). The inset shows a zoom into
  $\rho_\Omega(\Omega|K=2,0.4,T)$ for different values of $T$.}
\label{fig:frozenness-dist-k2-theory}
\end{figure}

Fig.~\ref{fig:frozenness-dist-k2} shows the distribution of $\Omega$
for simulated quenched RBNs with $K=2$ for different values of $p$,
compared with the result of the numerical evaluation of
Eq.~(\ref{eq:pq_allk}) as described above. There is a very good
agreement between the two types of results. The peaks correspond to
prominent values of the frozenness.  The rightmost peak is always due
to the constant functions, but large frozenness values are also
obtained for other functions. For instance, $f_1$ assumes the value 0
whenever both inputs are different.  If both inputs have $q=1/2$
(i.e., $\Omega=0$), the value of $q_{f=1}$ is $(1-2p)1/4+p$ and
$\Omega = ((1-2p)/2 + 2p - 1)^2$. This is the second main peak of
$\rho_\Omega(\Omega|2,0.4,10^4)$ (counted from the right end). For
smaller values of $p$, the peaks are not discernible, and a broad
continuum appears, with a distribution that follows a power-law with
an exponent $\simeq 0.6$ as $\Omega\to 0$. As $T$ becomes larger, it
is expected that the continuous regions become more and more
discontinuous, as can be seen in
Fig.~\ref{fig:frozenness-dist-k2-theory}, which shows the theoretical
prediction for larger times $T$. Moreover, when the resolution is
increased (see inset), it can be seen that peak-like regions which
appear like fluctuations around a single value of $\Omega$, are
in fact composed of sharper peaks, which themselves are composed of
other peaks, building a fractal structure.

Another distinguishing feature seen in Fig.~\ref{fig:frozenness-dist-k2} is a
sharp transition at $p=0$, where the only two possible values of $q$ are $0$ and
$1$, both of which amount to $\Omega=1$, leading to the variance
$\sigma_\Omega^2 = 0$. For $p>0$ this abruptly changes, and a wide range of
values of $q$ are possible, which discontinuously leads to
$\sigma_\Omega^2>0$. There is no such discontinuous transition for other $K$
values, but a continuous one (see following section), which makes the case $K=2$
special.

The function $\langle \Omega \rangle$ is obtained by performing the integral
 $\int \Omega \cdot \rho_\Omega(\Omega|K=2,p,T) d\Omega$. As can be seen in Figure
 \ref{fig:frozenness-vs-noise-better}, our calculation of this
 function agrees well with the results of computer simulations.

\subsection{$K>2$}\label{sec:kg2}

For larger values of $K$, numerical solutions of
Eq.~(\ref{eq:pq_allk}) become progressively more elaborate. We did not
pursue the task of writing the expressions of 256 function
$q_f(\{q^{(i)}\})$ for $K=3$ or of $65,536$ functions for $K=4$.
Instead, we perform in the following an approximation that is good for
a large number of inputs per node.

When $K$ is large, the vast majority of Boolean functions have the
output $1$ for approximately half the input combinations. This means
that almost all nodes have at their inputs $q$ values close to $1/2$.
We therefore make the assumption that the input values to each node
are $1$ and $0$ with probability $1/2$, independently from each other.
This means that for any given function all input combinations are
equally probable. It then follows immediately that the possible $q$
values are identical to the possible fractions of output values 1 in
the truth table of a Boolean function, and that the probability for a
given $q$ value is
\begin{equation}
p_q(q|K,p=0)=2^{-M} {M\choose qM}\, .
\end{equation}
Here, we have defined $M=2^K$, and the possible $q$ values are thus
multiples of $1/M$.

In the presence of noise, each output value is inverted with
probability $p$, implying that $q$ is changed to $q' =
q(1-p)+p(1-q) = (1-2p)q + p$. We therefore have
\begin{equation}
 \rho_q(q|K,p) = \frac{M}{1-2p} p_q\left(\left.\frac{q-p}{1-2p}\right|K,p=0\right)\, .
\end{equation}
For the frozenness $\Omega$ we obtain the distribution
\begin{equation}
\rho_\Omega(\Omega|K,p) = \frac{M}{2(1-2p)\sqrt{\Omega}}
p_q\left(\left.\frac{\sqrt{\Omega}+1-p}{2(1-2p)}\right|K,p\right)\, .
\end{equation}

Finally, one needs to take into account the effect of fluctuations, exactly as was done
for the previous cases,
\begin{equation}\label{eq:pq_largek}
  \rho_q(q|p,K,T) = \int_0^1\rho_q(q'|p,K)\tilde\rho_{q}(q|T,q')dq',
\end{equation}
where $\tilde\rho_{q}(q|T,q')$ is given by Eq.~(\ref{eq:q_T}).

\begin{figure}[h]
\includegraphics*[width=\columnwidth]{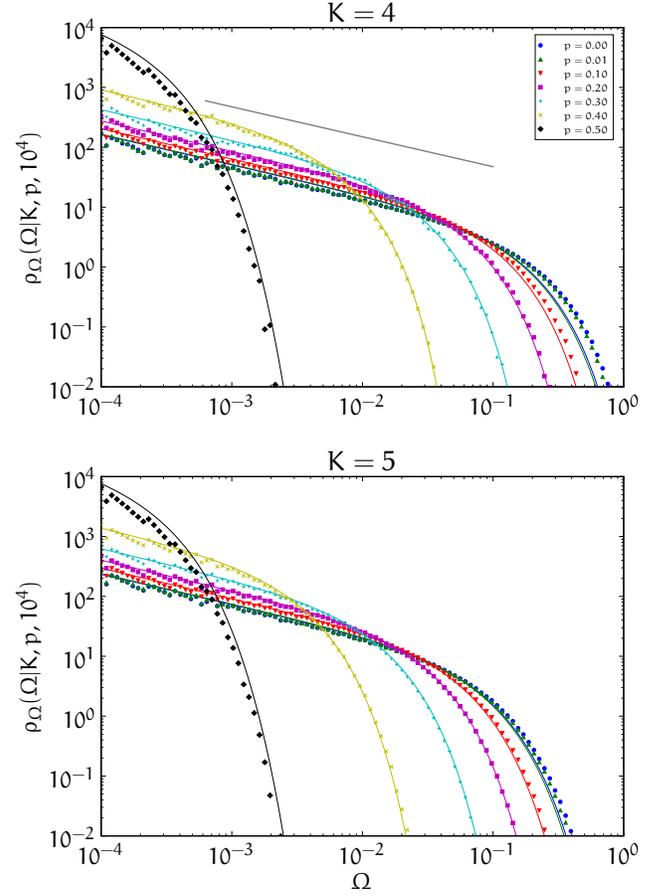}
\caption{(Color online) Distribution of the frozenness of nodes as function of
  noise strength for RBN of size $N=10^4$ for different values of $K$. Each
  curve was obtained by averaging the results for 100 different network
  realizations. The solid lines correspond to
  Eq.~\ref{eq:pq_largek}. \label{fig:frozenness-dist}}
\end{figure}

Fig.~\ref{fig:frozenness-dist} shows the distributions of frozenness for $K=4$
and $5$. In contrast to the cases $K=1$ and $K=2$, the peaks are less
pronounced, and are hardly visible. For $K=3$ (not shown) there are some peaks
which are still visible, specially for high values of $p$. Therefore, the
high-$K$ approximation Eq~(\ref{eq:pq_largek}) is very good already for $K=4$.
Both distributions show the same power-law decay
$\rho_\Omega(\Omega|K,p)\sim\Omega^{-1/2}$. This is simply due to the fact that
for $\Omega\to0$ ($q\to1/2$) the shape of $\rho_q(q|K,p)$ is essentially flat,
and thus $\rho_\Omega(\Omega|K,p) \sim dq/d\Omega = \Omega^{-1/2}/2$.

The quality of our approximation can also be assessed by comparing the
analytical prediction for $\left<\Omega\right>$ with the computer simulations
(Fig.~\ref{fig:frozenness-vs-noise-better}). We expect that the approximation
becomes even better for larger $K$.

\section{Conclusion}\label{sec:conclusion}

We have investigated the effect of thermal noise on RBNs by evaluating two order
parameters, the long time average of the Hamming distance between two networks
and the average frozenness of the network. While for $K=1$ and $K=2$ the average
Hamming distance increases continuously from 0 to 1/2 as $p$ increases from 0 to
1/2, it has a jump at $p=0$ for $K>2$.  These findings are well reproduced by
the annealed approximation, and they are a consequence of the transition from a
frozen to a ``chaotic'' phase in the deterministic system. In the chaotic phase
(occuring for $K>2$), initially nearby trajectories become eventually
uncorrelated. The smooth increase of the Hamming distance towards the value 1/2
is compatible with what was found in~\cite{miranda_noise_1989,
  qu_numerical_2002} for small networks.

The analysis of the average frozenness of the network required more
sophisticated calculations than the annealed approximation, and revealed
intricate details of the network dynamics. For all values of $K$ the probability
distribution of frozenness is a set of delta peaks. For $K=1$, these peaks can
be obtained by considering the distance of nodes to nodes with constant
functions. For $K>1$, the analysis becomes a lot more elaborate, due to the
large number of Boolean functions and the resulting vast number of possible
combinations of frozenness values for the inputs of each function. We explained
the general method, and performed the actual numerical evaluation for the case
$K=2$. The delta peaks show a fractal structure, which emerges from the iterated
recursion relation for the possible frozenness values. The variance of the
frozenness distribution changes continuously with $p$ for all $K \neq 2$, but
for $K=2$ it has a jump at $p=0$, where the variance changes discontinuously
from $0$ to a value larger than zero.  For larger values of $K$, the delta peaks
are so close to each other that the frozenness distribution appears continuous,
and in this limit we succeeded in performing an approximate analytical
calculation.

We do not find a phase transition at finite noise strength, in contrast to
\cite{huepe_dynamical_2002}, where Boolean networks with threshold functions
following a majority rule were used. Such a system undergoes a second order
phase transition from an ordered ``ferromagnetic'' phase, where all nodes assume
the same value for the majority of time, to a disordered phase, where the nodes
assume both states equally often. The presence of an ordered phase is a direct
consequence of the majority rule, and this transition is similar to that in a
network of Ising
spins~\cite{indekeu_special_2004,aleksiejuk_ferromagnetic_2002}.  The order
parameter in~\cite{huepe_dynamical_2002} was defined as the average
``alignment'' $s=\left|\left<1-2\sigma\right>\right|$, which is $1$ if all nodes
are in the same state.  The order parameter $s$ is only meaningful in systems
where the system is ordered in the absence of noise, and where the symmetry
between the states with values $0$ and $1$ is broken, as in ferromagnetic spin
systems.  Otherwise, $\left<\Omega\right>$ is a better order parameter, because
it captures disordered frozen phases, such as for $K=1$ in RBNs.  Of course, a
phase transition in the value of $s$ is always accompanied by a phase transition
in the value of $\left<\Omega\right>$. The opposite is not always true.

It is to be expected that real networks show some kind of robustness to noise,
since they must be able to carry out their function in a noisy environment. As
the results of this work show, only for RBNs with $K=1$ do the order parameters
change slowly as noise is switched on.  RBNs with $K>1$ fail to exhibit
robustness to noise, which is hardly surprising given the random wiring of the
system and the random choice of functions.  It will therefore be interesting to
extend the present study to networks with a more restricted set of functions
with more biological relevance, such as threshold~\cite{szejka_phase_2008} or
canalizing functions~\cite{moreira_canalizing_2005,kauffman_genetic_2004}.  At
least for some sets of functions, one should expect a phase transition at a
finite noise strength, similar to the transition seen
in~\cite{huepe_dynamical_2002}. The survival of the ``ordered'' phase up to a
certain noise strength can be viewed as a certain type of robustness.

It remains to be seen how other network topologies~\cite{iguchi_boolean_2007}
and the incorporation of redundancy~\cite{gershenson_role_2006} change a
network's response to noise. In~\cite{iguchi_boolean_2007}, it was shown that a
scale-free input distribution changes the average number and length of
attractors. In~\cite{gershenson_role_2006}, redundancy was introduced as
functional duplications of nodes in the network, which resulted in greater
robustness against random mutations of the update functions. In both papers,
only deterministic dynamics were considered. The effects of these (or other more
general) topological and functional characteristics may strongly alter the
response of a network to thermal noise. Finding the general conditions required
for reliable dynamics in a stochastic environment will be an important step
towards a deeper understanding of the dynamical features of real networks.

We acknowledge the support of this work by the Humboldt Foundation.

\bibliography{bib}

\begin{thebibliography}{27}
\expandafter\ifx\csname natexlab\endcsname\relax\def\natexlab#1{#1}\fi
\expandafter\ifx\csname bibnamefont\endcsname\relax
  \def\bibnamefont#1{#1}\fi
\expandafter\ifx\csname bibfnamefont\endcsname\relax
  \def\bibfnamefont#1{#1}\fi
\expandafter\ifx\csname citenamefont\endcsname\relax
  \def\citenamefont#1{#1}\fi
\expandafter\ifx\csname url\endcsname\relax
  \def\url#1{\texttt{#1}}\fi
\expandafter\ifx\csname urlprefix\endcsname\relax\def\urlprefix{URL }\fi
\providecommand{\bibinfo}[2]{#2}
\providecommand{\eprint}[2][]{\url{#2}}

\bibitem[{\citenamefont{Kauffman}(1969)}]{kauffman_metabolic_1969}
\bibinfo{author}{\bibfnamefont{S.~A.} \bibnamefont{Kauffman}},
  \bibinfo{journal}{J. Theor. Biol.} \textbf{\bibinfo{volume}{22}},
  \bibinfo{pages}{437} (\bibinfo{year}{1969}).

\bibitem[{\citenamefont{Drossel}(2008)}]{drossel_random_2008}
\bibinfo{author}{\bibfnamefont{B.}~\bibnamefont{Drossel}},
  \emph{\bibinfo{title}{Reviews of Nonlinear Dynamics and Complexity}}
  (\bibinfo{publisher}{Wiley}, \bibinfo{year}{2008}), vol.~\bibinfo{volume}{1},
  ISBN \bibinfo{isbn}{3527407294}.

\bibitem[{\citenamefont{Rosen-Zvi et~al.}(2001)\citenamefont{Rosen-Zvi, Engel,
  and Kanter}}]{rosen-zvi_multilayer_2001}
\bibinfo{author}{\bibfnamefont{M.}~\bibnamefont{Rosen-Zvi}},
  \bibinfo{author}{\bibfnamefont{A.}~\bibnamefont{Engel}}, \bibnamefont{and}
  \bibinfo{author}{\bibfnamefont{I.}~\bibnamefont{Kanter}},
  \bibinfo{journal}{Phys. Rev. Lett.} \textbf{\bibinfo{volume}{87}},
  \bibinfo{pages}{078101} (\bibinfo{year}{2001}).

\bibitem[{\citenamefont{Moreira et~al.}(2004)\citenamefont{Moreira, Mathur,
  Diermeier, and Amaral}}]{moreira_efficient_2004}
\bibinfo{author}{\bibfnamefont{A.~A.} \bibnamefont{Moreira}},
  \bibinfo{author}{\bibfnamefont{A.}~\bibnamefont{Mathur}},
  \bibinfo{author}{\bibfnamefont{D.}~\bibnamefont{Diermeier}},
  \bibnamefont{and} \bibinfo{author}{\bibfnamefont{L.~A.~N.}
  \bibnamefont{Amaral}}, \bibinfo{journal}{Proc. Nat. Ac. Sci.}
  \textbf{\bibinfo{volume}{101}}, \bibinfo{pages}{12085}
  (\bibinfo{year}{2004}).

\bibitem[{\citenamefont{Lagomarsino et~al.}(2005)\citenamefont{Lagomarsino,
  Jona, and Bassetti}}]{lagomarsino_logic_2005}
\bibinfo{author}{\bibfnamefont{M.~C.} \bibnamefont{Lagomarsino}},
  \bibinfo{author}{\bibfnamefont{P.}~\bibnamefont{Jona}}, \bibnamefont{and}
  \bibinfo{author}{\bibfnamefont{B.}~\bibnamefont{Bassetti}},
  \bibinfo{journal}{Phys. Rev. Lett.} \textbf{\bibinfo{volume}{95}},
  \bibinfo{pages}{158701} (\bibinfo{year}{2005}).

\bibitem[{\citenamefont{Bornholdt}(2005)}]{bornholdt_systems_2005}
\bibinfo{author}{\bibfnamefont{S.}~\bibnamefont{Bornholdt}},
  \bibinfo{journal}{Science} \textbf{\bibinfo{volume}{310}},
  \bibinfo{pages}{449} (\bibinfo{year}{2005}).

\bibitem[{\citenamefont{Greil and Drossel}(2005)}]{greil_dynamics_2005}
\bibinfo{author}{\bibfnamefont{F.}~\bibnamefont{Greil}} \bibnamefont{and}
  \bibinfo{author}{\bibfnamefont{B.}~\bibnamefont{Drossel}},
  \bibinfo{journal}{Phys. Rev. Lett.} \textbf{\bibinfo{volume}{95}},
  \bibinfo{pages}{048701} (\bibinfo{year}{2005}).

\bibitem[{\citenamefont{Klemm and Bornholdt}(2005)}]{klemm_stable_2005}
\bibinfo{author}{\bibfnamefont{K.}~\bibnamefont{Klemm}} \bibnamefont{and}
  \bibinfo{author}{\bibfnamefont{S.}~\bibnamefont{Bornholdt}},
  \bibinfo{journal}{Phys. Rev. E} \textbf{\bibinfo{volume}{72}},
  \bibinfo{pages}{055101} (\bibinfo{year}{2005}).

\bibitem[{\citenamefont{Samuelsson and
  Troein}(2003)}]{samuelsson_superpolynomial_2003}
\bibinfo{author}{\bibfnamefont{B.}~\bibnamefont{Samuelsson}} \bibnamefont{and}
  \bibinfo{author}{\bibfnamefont{C.}~\bibnamefont{Troein}},
  \bibinfo{journal}{Phys. Rev. Lett.} \textbf{\bibinfo{volume}{90}},
  \bibinfo{pages}{098701} (\bibinfo{year}{2003}).

\bibitem[{\citenamefont{Drossel}(2005)}]{drossel_number_2005-1}
\bibinfo{author}{\bibfnamefont{B.}~\bibnamefont{Drossel}},
  \bibinfo{journal}{Phys. Rev. E} \textbf{\bibinfo{volume}{72}},
  \bibinfo{pages}{016110} (\bibinfo{year}{2005}).

\bibitem[{\citenamefont{Drossel et~al.}(2005)\citenamefont{Drossel, Mihaljev,
  and Greil}}]{drossel_number_2005}
\bibinfo{author}{\bibfnamefont{B.}~\bibnamefont{Drossel}},
  \bibinfo{author}{\bibfnamefont{T.}~\bibnamefont{Mihaljev}}, \bibnamefont{and}
  \bibinfo{author}{\bibfnamefont{F.}~\bibnamefont{Greil}},
  \bibinfo{journal}{Phys. Rev. Lett.} \textbf{\bibinfo{volume}{94}},
  \bibinfo{pages}{088701} (\bibinfo{year}{2005}).

\bibitem[{\citenamefont{Shmulevich et~al.}(2002)\citenamefont{Shmulevich,
  Dougherty, Kim, and Zhang}}]{shmulevich_probabilistic_2002}
\bibinfo{author}{\bibfnamefont{I.}~\bibnamefont{Shmulevich}},
  \bibinfo{author}{\bibfnamefont{E.~R.} \bibnamefont{Dougherty}},
  \bibinfo{author}{\bibfnamefont{S.}~\bibnamefont{Kim}}, \bibnamefont{and}
  \bibinfo{author}{\bibfnamefont{W.}~\bibnamefont{Zhang}},
  \bibinfo{journal}{Bioinformatics} \textbf{\bibinfo{volume}{18}},
  \bibinfo{pages}{261} (\bibinfo{year}{2002}).

\bibitem[{\citenamefont{Indekeu}(2004)}]{indekeu_special_2004}
\bibinfo{author}{\bibfnamefont{J.~O.} \bibnamefont{Indekeu}},
  \bibinfo{journal}{Physica A} \textbf{\bibinfo{volume}{333}},
  \bibinfo{pages}{461} (\bibinfo{year}{2004}).

\bibitem[{\citenamefont{Aleksiejuk et~al.}(2002)\citenamefont{Aleksiejuk,
  Holyst, and Stauffer}}]{aleksiejuk_ferromagnetic_2002}
\bibinfo{author}{\bibfnamefont{A.}~\bibnamefont{Aleksiejuk}},
  \bibinfo{author}{\bibfnamefont{J.~A.} \bibnamefont{Holyst}},
  \bibnamefont{and} \bibinfo{author}{\bibfnamefont{D.}~\bibnamefont{Stauffer}},
  \bibinfo{journal}{Physica A} \textbf{\bibinfo{volume}{310}},
  \bibinfo{pages}{260} (\bibinfo{year}{2002}).

\bibitem[{\citenamefont{McAdams and Arkin}(1997)}]{mcadams_stochastic_1997}
\bibinfo{author}{\bibfnamefont{H.~H.} \bibnamefont{McAdams}} \bibnamefont{and}
  \bibinfo{author}{\bibfnamefont{A.}~\bibnamefont{Arkin}},
  \bibinfo{journal}{Proc. Nat. Ac. Sci.} \textbf{\bibinfo{volume}{94}},
  \bibinfo{pages}{814} (\bibinfo{year}{1997}).

\bibitem[{\citenamefont{Fretter and Drossel}(2008)}]{fretter_response_2008}
\bibinfo{author}{\bibfnamefont{C.}~\bibnamefont{Fretter}} \bibnamefont{and}
  \bibinfo{author}{\bibfnamefont{B.}~\bibnamefont{Drossel}},
  \bibinfo{journal}{Eur. Phys. J. B} \textbf{\bibinfo{volume}{62}},
  \bibinfo{pages}{365} (\bibinfo{year}{2008}).

\bibitem[{\citenamefont{Miranda and Parga}(1989)}]{miranda_noise_1989}
\bibinfo{author}{\bibfnamefont{E.~N.} \bibnamefont{Miranda}} \bibnamefont{and}
  \bibinfo{author}{\bibfnamefont{N.}~\bibnamefont{Parga}},
  \bibinfo{journal}{Europhys. Lett.} \textbf{\bibinfo{volume}{10}},
  \bibinfo{pages}{293} (\bibinfo{year}{1989}).

\bibitem[{\citenamefont{Golinelli and Derrida}(1989)}]{golinelli_barrier_1989}
\bibinfo{author}{\bibfnamefont{O.}~\bibnamefont{Golinelli}} \bibnamefont{and}
  \bibinfo{author}{\bibfnamefont{B.}~\bibnamefont{Derrida}},
  \bibinfo{journal}{J. Phys} \textbf{\bibinfo{volume}{50}},
  \bibinfo{pages}{1587} (\bibinfo{year}{1989}).

\bibitem[{\citenamefont{Qu et~al.}(2002)\citenamefont{Qu, Aldana, and
  Kadanoff}}]{qu_numerical_2002}
\bibinfo{author}{\bibfnamefont{X.}~\bibnamefont{Qu}},
  \bibinfo{author}{\bibfnamefont{M.}~\bibnamefont{Aldana}}, \bibnamefont{and}
  \bibinfo{author}{\bibfnamefont{L.~P.} \bibnamefont{Kadanoff}},
  \bibinfo{journal}{J. Stat. Phys.} \textbf{\bibinfo{volume}{109}},
  \bibinfo{pages}{967} (\bibinfo{year}{2002}).

\bibitem[{\citenamefont{Huepe and
  Aldana-Gonz{\'a}lez}(2002)}]{huepe_dynamical_2002}
\bibinfo{author}{\bibnamefont{Huepe}} \bibnamefont{and}
  \bibinfo{author}{\bibnamefont{Aldana-Gonz{\'a}lez}}, \bibinfo{journal}{J.
  Stat. Phys.} \textbf{\bibinfo{volume}{108}}, \bibinfo{pages}{527}
  (\bibinfo{year}{2002}).

\bibitem[{\citenamefont{Derrida and Pomeau}(1986)}]{derrida_random_1986}
\bibinfo{author}{\bibfnamefont{B.}~\bibnamefont{Derrida}} \bibnamefont{and}
  \bibinfo{author}{\bibfnamefont{Y.}~\bibnamefont{Pomeau}},
  \bibinfo{journal}{Europhys. Lett.} \textbf{\bibinfo{volume}{1}},
  \bibinfo{pages}{45} (\bibinfo{year}{1986}), ISSN \bibinfo{issn}{0295-5075}.

\bibitem[{\citenamefont{Kaufman and Drossel}(2005)}]{kaufman_properties_2005}
\bibinfo{author}{\bibnamefont{Kaufman}} \bibnamefont{and}
  \bibinfo{author}{\bibnamefont{Drossel}}, \bibinfo{journal}{Eur. Phys. J. B}
  \textbf{\bibinfo{volume}{43}}, \bibinfo{pages}{115} (\bibinfo{year}{2005}).

\bibitem[{\citenamefont{Szejka et~al.}(2008)\citenamefont{Szejka, Mihaljev, and
  Drossel}}]{szejka_phase_2008}
\bibinfo{author}{\bibfnamefont{A.}~\bibnamefont{Szejka}},
  \bibinfo{author}{\bibfnamefont{T.}~\bibnamefont{Mihaljev}}, \bibnamefont{and}
  \bibinfo{author}{\bibfnamefont{B.}~\bibnamefont{Drossel}},
  \bibinfo{journal}{New J. Phys.} \textbf{\bibinfo{volume}{10}},
  \bibinfo{pages}{063009} (\bibinfo{year}{2008}).

\bibitem[{\citenamefont{Moreira and Amaral}(2005)}]{moreira_canalizing_2005}
\bibinfo{author}{\bibfnamefont{A.~A.} \bibnamefont{Moreira}} \bibnamefont{and}
  \bibinfo{author}{\bibfnamefont{L.~A.~N.} \bibnamefont{Amaral}},
  \bibinfo{journal}{Phys. Rev. Lett.} \textbf{\bibinfo{volume}{94}},
  \bibinfo{pages}{218702} (\bibinfo{year}{2005}).

\bibitem[{\citenamefont{Kauffman et~al.}(2004)\citenamefont{Kauffman, Peterson,
  Samuelsson, and Troein}}]{kauffman_genetic_2004}
\bibinfo{author}{\bibfnamefont{S.}~\bibnamefont{Kauffman}},
  \bibinfo{author}{\bibfnamefont{C.}~\bibnamefont{Peterson}},
  \bibinfo{author}{\bibfnamefont{B.}~\bibnamefont{Samuelsson}},
  \bibnamefont{and} \bibinfo{author}{\bibfnamefont{C.}~\bibnamefont{Troein}},
  \bibinfo{journal}{Proc. Nat. Ac. Sci.} \textbf{\bibinfo{volume}{101}},
  \bibinfo{pages}{17102} (\bibinfo{year}{2004}).

\bibitem[{\citenamefont{Iguchi et~al.}(2007)\citenamefont{Iguchi, ichi
  Kinoshita, and Yamada}}]{iguchi_boolean_2007}
\bibinfo{author}{\bibfnamefont{K.}~\bibnamefont{Iguchi}},
  \bibinfo{author}{\bibfnamefont{S.}~\bibnamefont{ichi Kinoshita}},
  \bibnamefont{and} \bibinfo{author}{\bibfnamefont{H.~S.}
  \bibnamefont{Yamada}}, \bibinfo{journal}{J. Theor. Biol.}
  \textbf{\bibinfo{volume}{247}}, \bibinfo{pages}{138} (\bibinfo{year}{2007}).

\bibitem[{\citenamefont{Gershenson et~al.}(2006)\citenamefont{Gershenson,
  Kauffman, and Shmulevich}}]{gershenson_role_2006}
\bibinfo{author}{\bibfnamefont{C.}~\bibnamefont{Gershenson}},
  \bibinfo{author}{\bibfnamefont{S.~A.} \bibnamefont{Kauffman}},
  \bibnamefont{and}
  \bibinfo{author}{\bibfnamefont{I.}~\bibnamefont{Shmulevich}},
  \emph{\bibinfo{title}{Artificial Life X: Proceedings of the Tenth
  International Conference on the Simulation and Synthesis of Living Systems}}
  (\bibinfo{publisher}{The MIT Press}, \bibinfo{year}{2006}), ISBN
  \bibinfo{isbn}{0262681625}.

\end{thebibliography}

\end{document}